\documentclass[preprint,aps,nofootinbib]{revtex4}
\usepackage{amsfonts,amsmath,amssymb,amsthm}
\usepackage{latexsym}
\usepackage{bbm,bm}
\usepackage{graphicx}
\usepackage{tikz}
\usepackage{multirow}

\newcommand{\beq}{\begin{equation}}
\newcommand{\eeq}{\end{equation}}
\newcommand{\beqs}{\begin{eqnarray}}
\newcommand{\eeqs}{\end{eqnarray}}

\begin{document}

\title{A Simple Model of Superconductors: Insights from Free Fermion and Boson Gases}


                  
\author{ Mi-Ra Hwang$^1$, Eylee Jung$^1$, MuSeong Kim$^2$ and DaeKil Park$^{1,3}$\footnote{corresponding author, dkpark@kyungnam.ac.kr} }

\affiliation{ $^1$Department of Electronic Engineering, Kyungnam University, Changwon,
                 631-701, Korea    \\
                 $^2$Pharos iBio Co., Ltd. 
                Head Office: \#1408, 38, Heungan-daero 427beon-gil, Dongan-gu, Anyang, 14059, Korea \\
                $^3$Department of Physics, Kyungnam University, Changwon,
                  631-701, Korea }

\preprint{\bf{KMC-24-02}}
\begin{abstract}
Superconductors at temperatures below the critical temperature $T_c$
can be modeled as a mixture of Fermi and Bose gases, where the Fermi gas consists of conduction electrons and the Bose gas comprises Cooper pairs. 
This simple model enables the computation of the temperature dependence of $2 r(T) / N$,
where $N$ is the total number of conduction electrons and $r(T)$ is the number of Cooper pairs at temperature $T$.
Analyzing $2 r(T) / N$ across various superconductors may provide significant insights into the mechanisms behind high-temperature superconductivity, 
especially regarding coherence in Cooper pairs.
\end{abstract}

\maketitle
A Fermi gas is an idealized model for an ensemble of many non-interacting fermions. 
This model is useful to explore the electronic and thermal properties of the certain systems with many fermions such as the behavior of charge carries in a metal \cite{ashcroft}, nucleons in an atomic nucleus, 
neutrons in a neutron star, and electrons in a white dwarf.
Since spin-1/2 fermions obey the Fermi-Dirac statistics, the Fermi energy, the highest energy when the external temperature $T$ is zero, can be easily computed as follows:
\begin{eqnarray}
\label{fermi-1}
E_F = \left\{                    \begin{array}{cc}
                                     \frac{\pi \hbar^2}{m} \frac{N}{S}   &    \mbox{for} \hspace{.3cm}  d = 2    \\
                                     \frac{\hbar^2}{2 m} \left( \frac{3 \pi^2 N}{V} \right)^{2/3}     &     \mbox{for} \hspace{.3cm}  d=3,
                                       \end{array}                       \right.
\end{eqnarray}
where $d$ is a dimension of the system and $m$ is a mass of the individual spin-1/2 particle. In Eq. (\ref{fermi-1}) $N$, $S$, and $V$ are the total number of the fermions, surface area, and volume of the system respectively.
Thus, if we know the concentration $n = N / S$ or $N / V$, we can compute $E_F$ by applying Eq. (\ref{fermi-1}). 
Typical values of $E_F$ are $2 \sim 10$ eV for metals, $0.3$ MeV for white dwarfs, and $30$ MeV for nucleus.

If we do not know the total number of the spin-1/2 particles, it is possible to estimate $N$ by measuring $E_F$ experimentally as follows:
\begin{eqnarray}
\label{fermi-2}
N = \left\{                    \begin{array}{cc}
                                 \frac{m S}{\pi \hbar^2} E_F       &    \mbox{for} \hspace{.3cm}  d = 2    \\
                                 \frac{V}{3 \pi^2 \hbar^3} \left( 2 m E_F \right)^{3/2}       &     \mbox{for} \hspace{.3cm}  d=3.
                                       \end{array}                       \right.
\end{eqnarray}
For the case of metal one can measure $E_F$ by various methods such as the anaylsis of thermionic emission and diffusion\cite{efermi-1},  angle-resolved photoemission spectroscopy (ARPES) \cite{efermi-2}, and
ultraviolet photoelectron spectroscopy (UPS) \cite{efermi-3}. Therefore, using these experimental techniques and Eq. (\ref{fermi-2}) one can estimate $N$. 

In Ref. \cite{cooper} it was suggested that in the superconductor two electrons near the Fermi surface form a bound state by the attractive force mediated by the phonon. This pair is called ``Cooper pair''.
The formation of the Cooper pair implies the energy gap $\Delta (T)$ in the energy spectrum of superconductor. The energy gap is roughly dissolution energy, which is necessary to break the pair into the two free conduction electrons.
The gap $\Delta (T)$ exhibits a decreasing behavior with increasing the external temperature and eventually, goes to 
zero at the critical temperature $T_c$. The microscopic theory of superconductivity was proposed in Ref. \cite{bcs-1,bcs-2}, which is so called the BCS theory. 
The BCS theory explains many phenomena in the superconductor. One of them is that when the spin state of each Cooper pair is singlet,  the ratio between $\Delta_0 = \Delta(0)$ and $T_c$ is given as a form:
\begin{equation}
\label{ratio}
\frac{\Delta_0}{k_B T_c} = \frac{2 \pi}{e^{\gamma}} \approx 3.53
\end{equation}
where $\gamma = 0.5772\dots$ is a Euler constant. 

\begin{center}
Table I: $T_c$ and $\Delta_0$ for various superconductors
\vspace{0.2cm}
\tabcolsep=0.3cm

\begin{tabular}{c|c|c|c}\hline
               &   Material   &   $T_c, K$    &   $\Delta_0 / k_B T_c$       \\     \hline
               &    Hf       &    $0.13$       &   $3.9$                                \\       \cline{2-4}
               &   Cd       &    $0.52$       &   $3.2$                                 \\       \cline{2-4}
               &   Zn       &    $0.85$       &   $3.2$                                  \\       \cline{2-4}
               &  Al         &    $1.2$        &    $3.4$                                  \\       \cline{2-4}
         Low-$T_c$      &  In         &    $3.4$        &    $3.6$                                   \\       \cline{2-4}
               &  Hg       &     $4.2$       &     $4.6$                                    \\       \cline{2-4}
               &  Pb       &     $7.2$       &     $4.3$                                    \\       \cline{2-4}
               &  Nb       &     $9.3$       &     $3.8$                                    \\       \cline{2-4}
               & $\mbox{Ba}_{0.6}\mbox{K}_{0.4}\mbox{BiO}_3$     &   $18.5$    &    $3.7$               \\       \cline{2-4}
               & $\mbox{K}_3 \mbox{C}_{60}$    &     $19$      &    $3.6$                   \\   \hline
              &  $\mbox{YBa}_2\mbox{Cu}_3\mbox{O}_{7-\delta}$    &     $87$     &    $4.0$           \\       \cline{2-4}  
               &   $\mbox{Bi}_2\mbox{Sr}_2\mbox{Ca}_2\mbox{Cu}_3\mbox{O}_{10}$     &     $108$     &   $5.7$   \\       \cline{2-4}
        High-$T_c$       &  $\mbox{Tl}_2\mbox{Ba}_2 \mbox{CaCu}_2\mbox{O}_8$   &    $112$      &    $4.5$                \\       \cline{2-4}
               &  $\mbox{Tl}_2\mbox{Ba}_2 \mbox{Ca}_2\mbox{Cu}_3\mbox{O}_{10}$   &   $105$    &   $3.1$    \\       \cline{2-4}  
               &  $\mbox{HgBa}_2\mbox{Ca}_2\mbox{Cu}_3\mbox{O}_{8}$   &  $131$      &    $4.3$                      \\    \hline

\end{tabular}

\end{center}
The critical temperature $T_c$ and $\Delta_0 / k_B T_c$ is summarized for various superconductors in Table I \cite{super95}.
It is shown that Eq. (\ref{ratio}) roughly coincides with the experimental data for low-$T_c$ superconductors.
However, $\Delta_0 / k_B T_c$ for high-$T_c$ superconductors seems to be larger than $3.53$.
Therefore, we may need a new theory of superconductivity to explain the high-$T_c$ superconductivity.
Table I shows that $\Delta_0$ for the high-$T_c$ superconductor is larger than that for the low-$T_c$ superconductor. 
This fact indicates that the attractive force between the two electrons in the Cooper pair becomes stronger.
Thus, the phonon-mediated scenario for the Cooper pair should be modified. Many proposals were suggested in the context of the condensed matter theory\cite{carlson02}.
Another scenario was suggested in the context of gravity theory  by making use of AdS$_{d+1}$/CFT$_d$ correspondence \cite{ads_cft}.
Few years ago AdS/CFT was used to explore quark confinement\cite{rey-98}, quark-monopole interaction\cite{park01-1}, and Gross-Ooguri phase transition\cite{hungsoo01}.
More recently, it was also applied to the Hall effect\cite{Hartnoll:2007ai},  Nernst effects\cite{Hartnoll:2007ih,Hartnoll:2007ip,Hartnoll:2008hs}, and strange metal\cite{strange}.

Application of the AdS/CFT to the superconductivity was initiated in Ref.\cite{hart08} as a name of holographic superconductivity.
The ratio $\Delta_0 / k_B T_c$ in the holographic scenario was investigated in Ref.\cite{hart08,horo-78-1,gregory09}. Using simple black hole solution in AdS space, the ratio is given by 
\begin{equation}
\label{ratio2}
\frac{\Delta_0}{k_B T_c} \approx 8.2.
\end{equation}
Eq. (\ref{ratio2}) seems to be too large than that of the high-$T_c$ superconductors given in Table I. This fact may indicate that the holographic superconductivity is different from the 
conventional superconductivity arising in the condensed matter theory. Regardless of the fact I guess that holographic superconductivity may give useful insights in the future 
for developing a theory of the high-$T_c$ superconductor.

In this short note, we adopt a simplified picture where conduction electrons and Cooper pairs coexist in superconductors, viewed as ideal Fermi and Bose gases, respectively. Below $T_c$, some conduction electrons are converted into the Cooper pairs through an attractive interaction, typically mediated by phonons in conventional superconductors. These pairs act as bosons, allowing them to condense into a coherent state, a hallmark of the superconducting phase.

This simplified model captures essential aspects of superconductivity: the transformation of fermions into bosonic pairs and their condensation into a coherent state. Although the full microscopic treatment includes complex interactions, this approach offers an accessible framework for understanding the onset of superconductivity, especially when analyzing the temperature dependence of the Cooper pair population.

As $T$ decreases from $T_c$, number of Fermi (or boson) particles decreases (or increases). In this way the roles of the conduction electrons and Cooper pairs can be distinguished as follow.
With decreasing $T$, the Fermi energy $E_F$ of the Fermi gas become lowed, and hence the number of Cooper pairs increases, all of which condensate in the ground state of the Boson gas.
In summary, the Fermi gas supplies the Cooper pairs, and all Cooper pairs play the role of condensation. 
We think this is the simplest picture of the superconductivity.

The advantage of this simple picture is that it allows us to compute the ratio between number of Cooper pairs  and total conduction electron  as follows.
As far as we know, the temperature dependence of the ratio has not been experimentally measured across various superconductors.

Let us consider a material ${\cal A}$, which is a superconductor at $T \leq T_c$. Let us assume that the total number of the conduction electrons is $N$. 
This can be easily estimated by measuring the mole of  ${\cal A}$ and number of conduction electron(s) per atom.
For fixed $T$ with $T \leq T_c$ let us assume that the number of Cooper pairs is $r$. Then, it is evident that the number of the conduction electrons is 
$N - 2 r$.
Applying (\ref{fermi-2}) it is easy to derive 
\begin{eqnarray}
\frac{2 r}{N} = \left\{                 \begin{array}{cc}
                                      1 - \frac{m S}{\pi \hbar^2 N} E_F (T)      &    \mbox{for} \hspace{.3cm}  d = 2    \\
                                      1 - \frac{V}{3 \pi^2 \hbar^3 N} \left(2 m E_F (T) \right)^{3/2}     &     \mbox{for} \hspace{.3cm}  d=3.
                                       \end{array}                       \right.
\end{eqnarray}
Since  $E_F (T)$ is a Fermi energy at the temperature $T$, it can be measured experimentally. 
In this way one can measure $2 r / N$ at the temperature $T$.
If we repeat the experiment with changing the external temperature $T$, we can estimate the $T$-dependence of $2 r / N$ for the material  ${\cal A}$.
If we repeat the same experiment with changing the material  ${\cal A}$,  ${\cal B}$, $\cdots$, we can estimate the $T$-dependence of $2 r / N$ for many materials

Now, let us discuss how $2 r / N$ exhibits with respect to $T$. 
We can conjecture easily that $2 r/ N$ decreases with increasing $T$, because coherence among Cooper pairs diminishes near $T_c$.
However, we guess that $2 r / N$ does not go to zero at $T = T_c$, because the superconducting properties need the collective motion of the Cooper pairs. 
\begin{figure}[ht!]
\begin{center}
\includegraphics[height=5.9cm]{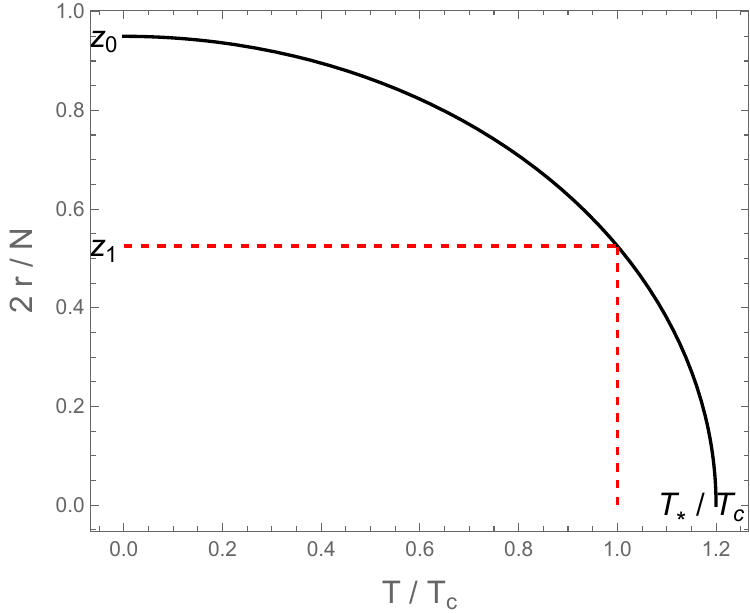} 
\includegraphics[height=5.9cm]{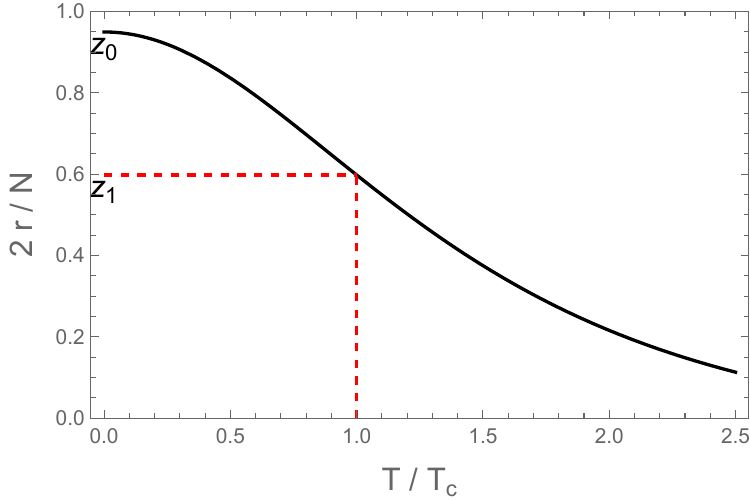}
\caption[fig1]{(Color online) The possible temperature dependence of $2 r / N$. In Fig. 1(a) $2 r / N$ decreases from $z_0 (< 1)$ to $z_1 (> 0)$ as $T$ increases from zero to $T_c$ and there exists $T_* (> T_c)$, where $2 r / N$ completely vanishes.
In Fig. 1(b) the $T$ dependence of $2 r / N$ approaches to zero as $T$ approaches to $\infty$. 

 }
\end{center}
\end{figure}
I guess the temperature dependence of $2 r / N$ would be either Fig. 1(a) or Fig. 1(b). In Fig. 1(a) $2 r / N$ decreases from $z_0 (< 1)$ to $z_1 (> 0)$ as $T$ increases from zero to $T_c$ and there exists $T_* (> T_c)$, where $2 r / N$ completely vanishes.
In Fig. 1(b) the $T$-dependence of $2 r / N$ exhibits a similar behavior with Fig. 1(a), but there is no $T_*$. This means that $2 r / N$ approaches to zero as $T$ approaches to $\infty$. 
It is worthwhile noting that the results of Fig. 1 are consistent with Ref. \cite{kjxu24}, where it was shown experimentally that the incoherent Cooper pairs exist even when $T > T_c$.

I guess the pattern of this figure is different in the low- and high-temperature superconductors. I think this difference may give important insights on the theory of the high-$T_c$ superconductor. 

At this stage many questions arise as follows. (1) In the Fig. 1(a)-type materials is $g = T_c / T_*$ universal or material-specific?
(2)If $g$ is material-specific, what properties of the superconductor determine $g$? (3) If some superconductors are Fig. 1(a)-type and others are Fig. 1(b)-type, what properties of the superconductors make this difference? 
Understanding these aspects could lead to deeper insights into the pairing mechanisms in superconductors.


\end{document}